\documentclass[journal]{IEEEtran}
\IEEEoverridecommandlockouts
\usepackage{cite}
\usepackage{fancyhdr,graphicx,amsmath,amssymb,amsfonts}
\usepackage{tabularx}
\usepackage{booktabs}
\usepackage[usestackEOL]{stackengine}
\edef\tmp{\the\baselineskip}
\setstackgap{L}{\tmp}
\usepackage{multirow}
\usepackage{makecell}
\usepackage{lscape}
\usepackage{algorithmic}
\usepackage{subcaption}
\usepackage{graphicx}
\usepackage[spaces,hyphens]{xurl}
\usepackage[colorlinks,allcolors=blue]{hyperref}
\usepackage{tikz}
\usepackage{textcomp}
\usepackage{xcolor}
\usepackage{soul}
\usepackage{pifont}
\usepackage{pgfplots}
\pgfplotsset{width=7cm,compat=1.9}
\def\BibTeX{{\rm B\kern-.05em{\sc i\kern-.025em b}\kern-.08em
    T\kern-.1667em\lower.7ex\hbox{E}\kern-.125emX}}

\usepackage{listings}

\colorlet{mygray}{black!30}
\colorlet{mygreen}{green!60!blue}
\colorlet{mymauve}{red!60!blue}

\newcommand{\cmark}{\ding{51}}
\newcommand{\xmark}{\ding{55}}

\begin{document}
\IEEEoverridecommandlockouts
\IEEEpubid{\makebox[\columnwidth]{1234-5-6789-0123-4/24/\$XX.00~\copyright2026 IEEE \hfill}
\hspace{\columnsep}\makebox[\columnwidth]{ }}
\IEEEpubidadjcol

\title{Software Vulnerability Detection Using a Lightweight Graph Neural Network}

\author{
    \IEEEauthorblockN{
        Miles Farmer,
        Ekincan Ufuktepe, \IEEEmembership{Member, IEEE},
        Anne Watson,
        Hialo Muniz Carvalho,
        Vadim Okun, \IEEEmembership{Member, IEEE},
        Zineb Maasaoui,
        Kannappan Palaniappan, \IEEEmembership{Senior Member, IEEE}
 	}

    \thanks{
        Miles Farmer, Anne Watson, Ekincan Ufuktepe, and Kannappan Palaniappan are with the Department of Electrical Engineering and Computer Science at the University of Missouri, MO, USA.
        (e-mails: mlfcmg@missouri.edu, akw8kk@missouri.edu, euh46@missouri.edu, pal@missouri.edu)

        Hialo Muniz Carvalho, Vadim Okun, and Zineb Maasaoui are with the Software and Systems Division at the National Institute of Standards and Technology, Gaithersburg, MD, USA
        (e-mails: hialo.munizcarvalho@nist.gov, vadim.okun@nist.gov, zineb.maasaoui@nist.gov)
    }
}

\maketitle
\markboth{IEEE Transactions on Software Engineering, Vol. X, No. Y, March 2026}
{Farmer, \MakeLowercase{\textit{(et al.)}}: Software Vulnerability Detection Using a Lightweight Graph Neural Network}

\bgroup
\def\thefootnote{}

\footnote{Disclaimer: Any opinions, findings, and conclusions or recommendations expressed in this publication are those of the authors and do not necessarily reflect the views of the U.S. Government or agency thereof. Commercial products may be identified in this document, but such identification does not imply recommendation or endorsement by NIST, nor that the products identified are necessarily the best available for the purpose.}
\egroup

\begin{abstract}
Large Language Models (LLMs) have emerged as a popular choice in vulnerability detection studies given their foundational capabilities, open source availability, and variety of models, but have limited scalability due to extensive
compute requirements. 
Using the natural graph relational structure of code, we show that our proposed graph neural network (GNN) based deep learning model VulGNN for vulnerability detection can achieve performance almost on par with LLMs, but is 100 times smaller in size and fast to retrain and customize. We describe the VulGNN architecture, ablation studies on components, learning rates, and generalizability to different code datasets. As a lightweight model for vulnerability analysis, VulGNN is efficient and deployable at the edge as part of real-world software development pipelines. 

\end{abstract}


\section{Introduction} \label{sec:intro}

\IEEEPARstart{C}{onventional} techniques have shown limitations in detecting realistic and diverse software code vulnerabilities \cite{ghaffarian2017software}\cite{zeng2020software}\cite{hanif2021rise}\cite{shuai2015software}, prompting researchers to pursue deep learning approaches. Systems based on deep learning provided notable improvements in detection performance compared to conventional methods.
However, recent studies have also highlighted the limitations of deep learning-based vulnerability detectors when applied to complex, real-world data \cite{chen2023diversevul}. In such studies, benchmark GNN-based detectors lag behind LLM-based detectors in terms of accuracy and other detection metrics, though maintain a significant computational performance advantage due to their smaller model size. This performance advantage is of particular note and prompts further investigation, as one of the primary environments for which vulnerability detectors are desired, continuous integration/continuous delivery (CI/CD) pipelines, is often limited in compute resources, only allocated enough to compile and test the software. LLMs of hundreds of millions or billions of parameters are not practical to host on such hardware, necessitating a more efficient solution.

Given the inherent underlying graph structure of code, it follows that deep models acting on graph-structured code data may provide improvements over models that act directly on source code. An LLM interpreting raw source code is forced to determine structural information internally through learned parameters, which may be inefficient in training and real-world inference. Meanwhile, a graph model offloads that responsibility to algorithmic preprocessing steps, thereby lowering the intensity of the model and potentially allowing for more efficient structure interpretation.

Despite this assumed advantage, recent studies still demonstrate a disparity in detection accuracy between GNNs and LLMs. This work investigated the continued feasibility of graph-based vulnerability detection approaches by implementing several key components of language models within a GNN architecture. Furthermore, we conducted several generalization studies to determine the real-world generalizability of our model, along with the impact of real-world versus synthetic data in the training process. Specifically, we investigated the following:

\textit{\textbf{RQ1: How do GNN architecture variations affect vulnerability detection performance?}} We compare VulGNN to existing GNN architectures and investigate the impact of variations in the VulGNN architecture on vulnerability detection performance.

\textit{\textbf{RQ2: How does VulGNN compare to LLMs in real-world vulnerability detection performance?}} We apply VulGNN to the DiverseVul \cite{chen2023diversevul} dataset and compare its detection performance to that of LLMs applied to the same data.


\textit{\textbf{RQ3: Does VulGNN's capability generalize within and across datasets?}} We apply VulGNN to the ``unseen projects'' configuration of DiverseVul to test the model's ability to generalize to entirely unseen codebases. Furthermore, we perform an ``unseen dataset'' experiment, wherein the model is trained on one dataset and tested on another from a different source, to ascertain cross-dataset generalizability.

\textit{\textbf{RQ4: What is the impact of training on synthetic versus real-world data?}} A natural extension of the cross-dataset generalizability study, we perform a stepped ablation study to determine the impact of synthetic versus real-world training data on real-world vulnerability detection performance.

This paper makes the following main contributions:
\begin{enumerate}
    \item We propose an efficient GNN design and provide a systematic evaluation against LLM baselines.
    \item We investigate the impact of key GNN architecture choices on vulnerability detection performance.
    \item We analyze VulGNN's vulnerability detection performance and generalizability by applying it to the DiverseVul\cite{chen2023diversevul} dataset and comparing its performance to that of LLMs.
    \item We perform a stepped ablation study to compare the efficacy of real-world, synthetic, and mixed training data in real-world unseen vulnerability detection.
    \item We publicly share our source code\footnote{\url{https://github.com/CIVA-Lab/VulGNN}} for the model and its preprocessing steps for reproducibility.
\end{enumerate}

The manuscript is organized as follows. In Section \ref{sec:background}, the background of different types of code graph representations is explained. We also mention the data representation approaches used in state-of-the-art graph and language-based detection models. Section \ref{sec:modelarch} describes the model architecture and baseline parameters used in the experiments which follow. In Section \ref{sec:case}, we compare our resulting model to existing approaches (addressing RQ2). In Section \ref{sec:discuss}, we interpret the meaning of the case study results and discuss their significance in the context of current approaches. Additionally, we note the potential limitations of the study. In Section \ref{sec:related}, a summary of related work on deep learning-based vulnerability detection is given. Finally, Section \ref{sec:conclusion} concludes the paper and mentions the future work in relation to our study.

\section{Background and Fundamentals} \label{sec:background}

\subsection{Code Graph Representations} \label{subsec:background_graph_representations}
This section provides some preliminary information on the types of code graph representations.

\subsubsection{Abstract Syntax Trees}\label{sec:ast}
An Abstract Syntax Tree (AST)~\cite{landin1966next,lucas1981formal} is a hierarchical structure representing the abstract syntactic structure of a source code. It captures the essential grammatical structure and semantics of a program, facilitating analysis, transformation, and optimization by abstracting away irrelevant details like comments. ASTs play a critical role in programming language processing and software development tools and techniques.

\subsubsection{Control-Flow Graphs}\label{sec:cfg}
A Control-Flow Graph (CFG)~\cite{allen1970control,prosser1959applications} models the flow of control within a program, representing execution paths and decision points. Nodes represent basic blocks of codes, and edges indicate possible control flow. CFGs are key to program analysis, optimization, and error detection.

\subsubsection{Program Dependency Graphs}\label{sec:pdg}
A Program Dependency Graph (PDG)~\cite{ferrante1987program} represents relationships among program entities, capturing data and control dependencies. Data Dependency Graphs (DDGs)~\cite{kuck1981dependence,ottenstein1978data} focus on the data flow, while Control Dependency Graphs (CDGs)~\cite{ferrante1983program,natour1988control} focus on execution order. PDGs help in understanding information flow and control flow in a program.

\subsubsection{Code Property Graphs}\label{sec:cpg}
Code Property Graphs (CPGs)~\cite{cpg} provide a unified representation of code by integrating structural information from ASTs, CFGs, and PDGs. This holistic approach aids in advanced analysis, such as vulnerability detection and code comprehension. To generate CPGs we used Joern\footnote{\url{https://github.com/joernio/joern}}, which has two versions of CPGs: CPG14 and CPG. CPG14 refers to schema version 1.4 of the code property graph, representing an earlier, stable layout of node and edge types used in Joern. Compared to newer CPG versions, CPG14 has fewer node types similar edge relationships, and limited support for modern language features or multi-language analysis. While CPG14 is ideal for legacy compatibility and reproducibility, the default (latest) CPG schema provides richer representations, more detailed semantic properties, and better support for evolving analysis workflows. This enhanced expressiveness and granularity in the modern CPG schema make it better fit for machine learning-based vulnerability detection, as it captures more contextual and semantic information that models can leverage for improved accuracy and generalization. Therefore, in this study, we have used the latest CPG version (not CPG14).

\subsection{Sentence Embedding Models}
In this study, we used tokenization to preprocess code language into streams of integers compatible with embedding. In particular, we used the StarCoder language model's tokenizer, as it was trained on a variety of code in multiple modern languages, including C.

StarCoder is a GPT-style Transformer-decoder LLM proposed by Li et al. \cite{li2023starcoder}, built on top of the StarCoderBase model and Google AI PaLM 2 \cite{chowdhery2023palm}, which was trained using 80+ different programming languages with an emphasis on Python. The StarCoder and the StarCoderBase models were trained using a multi-query attention \cite{shazeer2019fast} architecture and Fill-in-the-Middle \cite{bavarian2022efficient} objective. Its tokenizer is of a Byte Pair Encoding design.

\section{Deep Learning Model Architecture and Parameters} \label{sec:modelarch}
\begin{figure*}[t]
    \centering
    \includegraphics[width=0.9\textwidth]{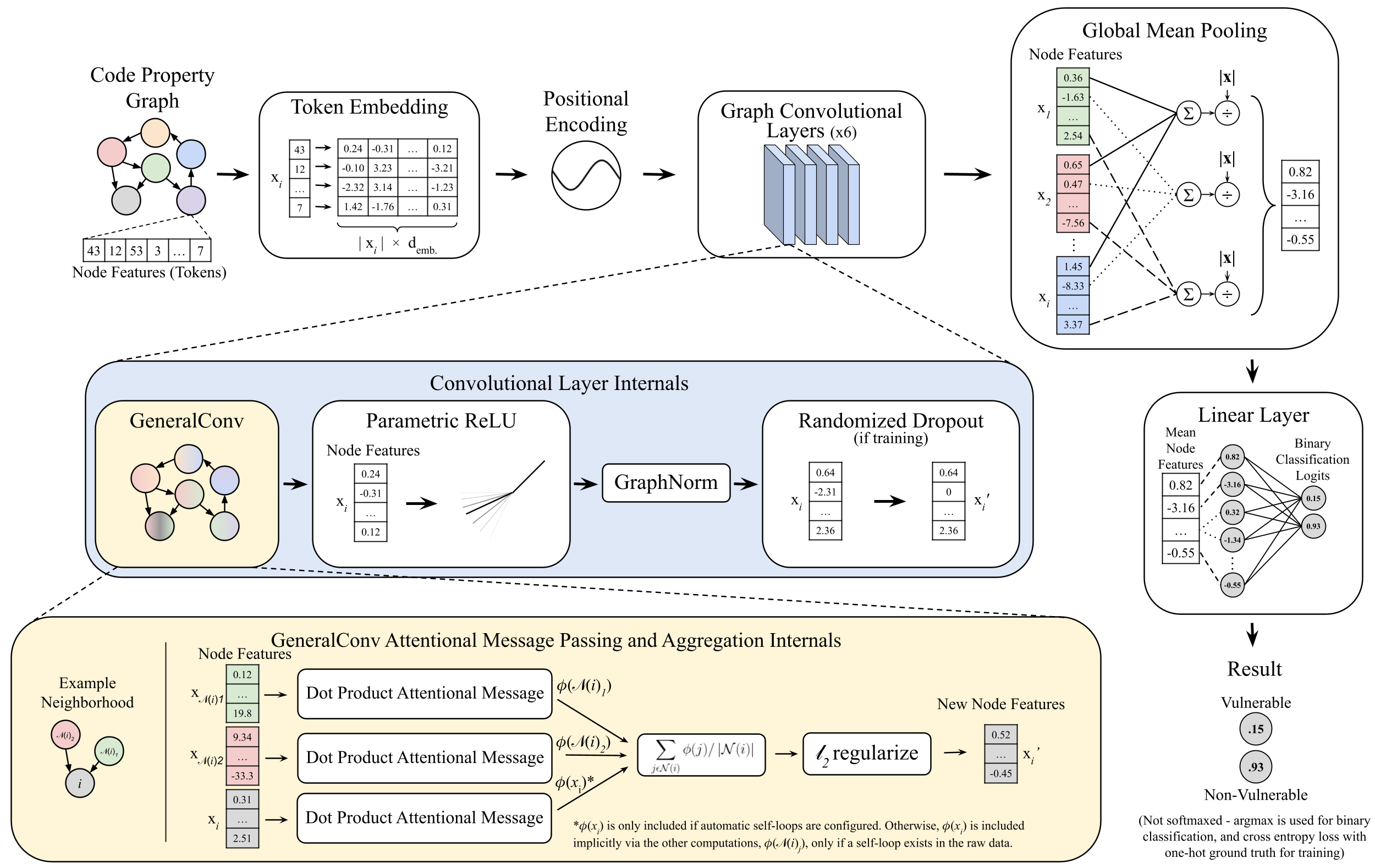}
    \caption{VulGNN graph convolutional deep architecture with preprocessing (including language tokenization) to generate the input Code Property Graph (Row 1), the internal details of the graph neural network layers (Row 2), and the implementation of the GeneralConv operator with graph attention (Row 3). The output from the GNN after Global Mean Pooling is passed to a single linear classification layer. VulGNN does not include a hidden MLP layer after pooling.}
    \label{fig:architecture}
\end{figure*}

The components of our VulGNN architecture are shown in Figure \ref{fig:architecture} and implemented using PyTorch and PyTorch Geometric \cite{pytorchgeometric}. 
Note that in this section and the paper overall, we use the terms graph neural network (GNN), graph convolutional network (GCN), and graph attention network (GAT) interchangeably. Since the VulGNN model described in this work is a GAT, the more general terms of GCN and GNN are applicable.

\subsection{Overview}
We propose \textsc{VulGNN}, a GNN designed for whole-graph binary classification of CPGs. The model uses attention-based message passing to capture dependencies between program elements and is configurable to operate with or without edge features. Nodes and edges are represented either as token sequences or as discrete types. Token sequences are embedded into a shared latent space and enriched with sinusoidal positional encodings. The architecture is composed of stacked convolutional blocks with attention, followed by graph-level pooling and a binary classification head. In our source code, the hyperparameters are controlled through a unified configuration (\texttt{ModelConfig}), enabling systematic ablations.

\subsection{Graph Representation}
Let $G=(V,E)$ denote a graph with $|V|=N$ nodes and $|E|$ edges. Each node $i \in V$ provides either:
\begin{itemize}
    \item a token sequence $X_i \in \mathbb{N}^{L_{\text{node}}}$ (default $L_{\text{node}}=8$), or
    \item a discrete node type $t_i \in \{1,\dots,|\mathcal{T}|\}$, where $|\mathcal{T}|=44$ is the number of node types in the CPG specification.
\end{itemize}
Similarly, each edge $(i,j)\in E$ provides either:
\begin{itemize}
    \item a token sequence $A_{ij} \in \mathbb{N}^{L_{\text{edge}}}$ (default $L_{\text{edge}}=16$), or
    \item a discrete edge type $r_{ij} \in \{1,\dots,|\mathcal{R}|\}$, where $|\mathcal{R}|=20$ corresponds to the set of CPG edge relations.
\end{itemize}

A batch of graphs is represented with a standard \texttt{batch} vector as in PyTorch Geometric~\cite{pytorchgeometric}, and graph-level readout is performed via mean pooling.

\subsection{Input Encoding}
A single embedding matrix $E \in \mathbb{R}^{V \times d}$ (with $V=49{,}152$, $d=16$) is shared for node and edge tokens:
\[
Z_i = E[X_i] \in \mathbb{R}^{L_{\text{node}} \times d}, \qquad 
A_{ij}^{\text{emb}} = E[A_{ij}] \in \mathbb{R}^{L_{\text{edge}} \times d}.
\]
Each token sequence is enriched with sinusoidal positional encodings~\cite{vaswani2017attention} and flattened:

\[
\tilde{Z}_i = \mathrm{vec}(\mathrm{PE}(Z_i)) \in \mathbb{R}^{L_{\text{node}}d}, 
\]
\[
\qquad \tilde{A}_{ij} = \mathrm{vec}(\mathrm{PE}(A_{ij}^{\text{emb}})) \in \mathbb{R}^{L_{\text{edge}}d}.
\]

When using node or edge \emph{types} in addition to or instead of tokenized sequences, we embed discrete IDs into low-dimensional vectors. In edge-type variants, edges are assigned embeddings $E_{\text{type}}[r_{ij}] \in \mathbb{R}^{d_e}$ with $d_e=4$.

\subsection{Convolutional Blocks}\label{sec:convblocks}
The backbone of VulGNN consists of stacked \textbf{ConvGroup} blocks. Each block applies:
\begin{enumerate}
    \item \textbf{Message passing:} a \texttt{GeneralConv}~\cite{you2021design} layer with dot-product attention, mean aggregation, and optional edge attributes. The \texttt{GeneralConv} is a general-purpose, highly-customizable graph convolutional layer inspired by You et al.~\cite{you2021design} that features support for multiple aggregation schemes, additive or dot-product single- or multi-head attention, directed or undirected edges, and edge attributes. Its implementation in this work is configured with mean aggregation and dot-product attention.
    \item \textbf{Nonlinearity:} PReLU activation.
    \item \textbf{Normalization:} For normalization, we have used the GraphNorm \cite{graphnorm} operation, where \(\hat{h}_{i,j}\) is the \textit{j}-th feature of the \textit{i}-th node, is defined as
\begin{equation}
    \mathrm{GraphNorm}(\hat{h}_{i,j}) = \gamma_j \cdot \frac{\hat{h}_{i,j} - \alpha_j\cdot\mu_j}{\hat{\sigma}_j} + \beta_j
\end{equation}
where $\mu_j=\frac{\sum_{i=1}^n\hat{h}_{i,j}}{n}, 
\hat{\sigma}_j^2 = \frac{\sum_{i=1}^n(\hat{h}_{i,j}-\alpha_j\cdot\mu_j)^2} {n} + \epsilon$ ; $\gamma_j, \beta_j$ are learnable affine parameters; and \(\alpha_j\) is a learnable weight parameter.
The original implementation by Cai et al. \cite{graphnorm} and the implementation we use from PyTorch Geometric \cite{pytorchgeometric} adds a small constant \(\epsilon = 10^{-5}\) to the variance for numerical stability.
    \item \textbf{Regularization:} Dropout with probability $p=0.08$.
\end{enumerate}

Formally, if $h^{(\ell)}$ denotes the node representation at layer $\ell$,
\[
h^{(\ell+1)} = \mathrm{Dropout}\!\left(\mathrm{GraphNorm}\!\left(\mathrm{PReLU}\!\left(\mathcal{G}^{(\ell)}(h^{(\ell)}, E)\right)\right)\right),
\]
where $\mathcal{G}^{(\ell)}$ is the \texttt{GeneralConv} operator.

The first block maps $L_{\text{node}}d \to D$ (default $8 \times 16 \to 128$), while subsequent blocks maintain hidden width $D=128$. By default, the model stacks six such layers.

\subsection{Graph Readout and Classification Head}
After $L$ convolutional layers, node embeddings are aggregated with global mean pooling:
\begin{equation}
\bar{h} = \frac{1}{N}\sum_{i=1}^N h_i^{(L)}.
\end{equation}
A linear projection preserves hidden size $D$, followed by a final linear layer that outputs two logits for binary classification:
\begin{equation}
\hat{y} = W_2\,\sigma(W_1 \bar{h}) \in \mathbb{R}^2.
\end{equation}

\subsection{Training Objective}
The model is trained using a weighted binary cross-entropy with logits objective, to handle class imbalance in the training dataset.
Given a raw logit $x \in \mathbb{R}$ class score and a binary target label $y \in \{0,1\}$, and weight $w$, the per-sample binary cross-entropy loss is defined as
\begin{equation}
\ell(x, y) = - \left( w \cdot y \cdot \log(\sigma(x)) + (1-y) \cdot \log\big(1 - \sigma(x)\big) \right),
\end{equation}
where $\sigma(\cdot)$ is the sigmoid function:
\begin{equation}
\sigma(x) = \frac{1}{1 + e^{-x}}.
\end{equation}
To avoid numerical issues for large or small values of $x$, PyTorch uses an equivalent but stable formulation:
\begin{equation}
\ell(x, y) = \max(x, 0) - x \cdot y + \log\!\big(1 + e^{-|x|}\big).
\end{equation}


\begin{equation}
w_k = \frac{N_{samples}}{N_{classes} \cdot N_k}
\end{equation}

\noindent \textbf{Where:}
\begin{itemize}
    \item $w_k$ is the class weight for class $k$.
    \item $N_{samples}$ is the total number of samples in the dataset.
    \item $N_{classes}$ is the total number of unique classes found in the data.
    \item $N_k$ is the count of samples belonging to class $k$.
\end{itemize}
If a positive class weight $\text{pos\_weight} \in \mathbb{R}^{+}$ is provided, the contribution of positive examples is scaled accordingly:

\begin{equation}
\begin{aligned}
\ell(x, y) &= \max(-x, 0) + \big(1-y\big) \cdot x +  \\
& \big(1 + (\text{pos\_weight} - 1) \cdot y \big) \cdot \left( \log\!\big(1 + e^{-|x|}\big) \right).
\end{aligned}
\end{equation}

For a batch of $N$ samples, the final loss is the mean over all examples:
\begin{equation}
L(x, y) = \frac{1}{N} \sum_{i=1}^{N} \ell(x_i, y_i).
\end{equation}

The optimization is performed using Adaptive Moment (ADAM) estimation  with learning rate $10^{-3}$, decay rates beta1 (momentum) of $0.9$, and beta2 (variance) of $0.999$, batch size $400$, and $25$ epochs. 

\subsection{Model Variations}
VulGNN supports configuration of several key aspects of the network, controlled by \texttt{ModelVariations}. The options for each key aspect are:
\begin{itemize}
    \item \textbf{Node representation:} tokenized language nodes vs. discrete node types (both embedded).
    \item \textbf{Edge representation:} no edge features, embedded edge types, or embedded rich edge information.
    \item \textbf{Backbone:} homogeneous attention-based convolution (\texttt{GeneralConv}), heterogeneous attention-based convolution (\texttt{RGATConv}).
\end{itemize}
This modularity enables systematic ablation of the effects of node features, edge features, and edge types on downstream performance.

Data processing was conducted across 6 to 13 servers, depending on the size of dataset and server availability, each of which has dual AMD EPYC 7713 CPUs (for a total of 128 cores) and 490GB of RAM. Model training was conducted on a single Nvidia H100 (80GB) running on a server with dual Xeon Platinum 8470s (total of 108 cores) and 2002 GB of RAM - though, we generally only utilize 32 cores for data loading and 96GB of RAM to keep the dataset in memory for low latency. Both of these steps can be accomplished on much less performant hardware, but this leads to longer processing and training times that inhibit efficient iterative testing. Inference can be accomplished on significantly less extensive hardware - the model uses only 449MB of VRAM with float32 used throughout. This could easily be lessened with little to no impact on classification performance with bfloat16 quantization \cite{bfloat16performance}, while potentially improving classification speed on newer hardware optimized for bfloat16 operations.

\section{Case Study} \label{sec:case}

To address RQ2 and RQ3, we compare the VulGNN system with existing state-of-the-art vulnerability detection systems. To accomplish this, we apply VulGNN to the DiverseVul dataset \cite{chen2023diversevul}, allowing for direct performance comparisons to other detectors assessed in that work. 

\subsection{Datasets}

We use the dataset provided by the DiverseVul \cite{chen2023diversevul} paper to benchmark VulGNN against existing methods. We refer to DiverseVul dataset, which includes Previous (deduplicated Devign~\cite{zhou2019devign}, ReVeal~\cite{chakraborty2021deep}, BigVul~\cite{bigvul}, CrossVul~\cite{crossvul} and CVEFixes~\cite{cvefixes}) as described in~\cite{chen2023diversevul}, containing 523,956 functions, with 41,377 being vulnerable samples and the remaining 482,579 being non-vulnerable. The Software Assurance Reference Dataset (SARD/Juliet) created by NIST is a collection of programs with documented weaknesses in several different languages (C, C++, Java, PHP, C\#), containing more than $450,000$ test cases ranging from small pieces of code to production software~\cite{nistsard}\cite{nistsardwebsite}. The Juliet Test Suite is a subset of the SARD/Juliet, containing $64,099$ small synthetic test cases, originally created by the NSA's Center for Assured Software (CAS) specifically for use in testing static analysis tools. Our SARD/Juliet training set consists of 33,360 subset of cases with 12,303 vulnerable functions and 21,057 non-vulnerable functions.

\subsection{Training and Test Procedures} \label{subsec:case_study_training_proc}
Following the methodology of Chen et al.~\cite{chen2023diversevul}, we adopt the same dataset splitting strategy to ensure direct comparability. All experiments use deduplicated function-level samples, with an $80/10/10$ ratio applied to construct the training, validation, and test splits, respectively. The validation set is used exclusively to select the best model checkpoint, i.e., the epoch yielding the highest validation F1-score. This checkpoint is then applied to the unseen test set for final performance reporting. This procedure minimizes overfitting and ensures fair evaluation on strictly held-out samples.

Given True Positives (TP), False Positives (FP), True Negatives (TN), and False Negatives (FN), the performance metrics used in this paper are computed as follows:

\begin{equation}
     Accuracy = \frac{TP + TN}{TP + FP + TN + FN}
\end{equation}
\begin{equation}
     Precision = \frac{TP}{TP + FP}
\end{equation}
\begin{equation}
     Recall = \frac{TP}{TP + FN}
\end{equation}
\begin{equation}
     F1 = 2 * \frac{Precision * Recall}{Precision + Recall}
\end{equation}

\subsection{Test I: Randomly Chosen Sets} \label{subsec:experiment_prev_plus_div}

The first evaluation follows the random splitting protocol defined in DiverseVul. Here, the merged corpus (Previous + DiverseVul) is randomly partitioned into $80\%$ training, $10\%$ validation, and $10\%$ test, while ensuring that no overlapping functions appear across splits. This approach provides a standard benchmark for assessing detector performance under balanced and representative data partitions. Importantly, this configuration permits direct comparison with prior baselines, as the exact same splitting criteria are maintained. Moreover, we use the exact splits provided by the DiverseVul paper to ensure our training, validation, and test sets are identical to those used previously.


\subsection{Test II: Unseen Project Generalizability Test} \label{subsec:experiment_generalizability}
Beyond random splits, we replicate the more realistic project-based evaluation of DiverseVul~\cite{chen2023diversevul}. In this setting, entire projects are held out to construct the test set, ensuring that no functions from those projects appear in training or validation. Concretely, 95 projects are randomly chosen as unseen, while the remainder form the seen portion of the dataset. The seen subset is again divided into $90\%$ training and $10\%$ validation. This design probes a detector's ability to generalize across project boundaries, which is an essential capability for deployment in real-world CI/CD environments, where detectors must handle vulnerabilities in previously unseen codebases.

\subsection{Training Data Augmentation Tests}
To explore the impact of dataset composition, we also adapt the DiverseVul~\cite{chen2023diversevul} ablation-style experiments. In particular, subsets of the training data are constructed to evaluate (i) the effect of scaling down to smaller training pools of only synthetic data, such as SARD/Juliet, and (ii) the balance between vulnerable and non-vulnerable samples. These controlled variations allow us to measure how performance shifts under constrained training regimes, highlighting the degree to which the detector benefits from larger and more diverse data sources.

\subsubsection{Test III: Real-World Training Data Ablation} \label{subsubsec:experiment_generalizability_realworld_train}

\begin{figure*}
\begin{subfigure}[t]{0.48\textwidth}
\begin{tikzpicture}[scale=1.2]
\begin{axis}[
    xlabel={\% Real-World Dataset Added to Juliet},
    ylabel={Accuracy},
    xmin=-5, xmax=105,
    ymin=50, ymax=100,
    xtick={0,20,40,60,80,100},
    ytick={50, 60, 70, 80, 90, 100},
    ymajorgrids=true,
    grid style=dashed,
    axis y line*=left
]
\addplot[
    color=blue,
    mark=square,
    ]
    coordinates {
    (0, 61.94) (10, 89.86) (20, 89.83) (40, 89.81) (60, 89.23) (80, 90.23) (100, 89.38)
    }; \label{real_world_data_acc_plot}
\end{axis}

\begin{axis}[
    ylabel={F1 Score},
    xmin=-5, xmax=105,
    ymin=0, ymax=45,
    ytick={0, 10, 20, 30, 40, 50},
    legend pos=south east,
    axis y line*=right
]
\addlegendimage{/pgfplots/refstyle=real_world_data_acc_plot}\addlegendentry{Accuracy}

\addplot[
    color=red,
    mark=o,
    ]
    coordinates {
    (0, 14.17) (10, 30.86) (20, 34.99) (40, 38.58) (60, 39.13) (80, 39.13) (100, 40.27)
    };
    \addlegendentry{F1 Score}
\end{axis}
\end{tikzpicture}
\end{subfigure}
\begin{subfigure}[t]{0.48\textwidth}
\begin{tikzpicture}[scale=0.97]

\begin{axis}[
    xlabel={Epoch},
    ylabel={F1 Score},
    xmin=0, xmax=26,
    ymin=0, ymax=45,
    ytick={0, 10, 20, 30, 40, 50},
    grid style=dashed,
    ymajorgrids=true,
    legend pos=south east,
    width=8.5cm
]

\addplot[
    color=blue,
    mark=square*,
    ]
    coordinates {
    (1, 0) (2, 12.92) (3, 13.44) (4, 13.5) (5, 13.63) (6, 13.12) (7, 13.1) (8, 12.93) (9, 12.98) (10, 13.08) (11, 13.44) (12, 13.34) (13, 13.37) (14, 13.22) (15, 13.01) (16, 13.37) (17, 13.05) (18, 12.89) (19, 12.81) (20, 12.73) (21, 12.51) (22, 13.33) (23, 12.84) (24, 13.11) (25, 13.09)
    }; \addlegendentry{0\%}

\addplot[
    color=red,
    mark=*,
    ]
    coordinates {
    (1, 1.49) (2, 2.37) (3, 6.29) (4, 12.59) (5, 17.49) (6, 26.81) (7, 23.48) (8, 29.31) (9, 28.44) (10, 25.71) (11, 26.41) (12, 31.56) (13, 29.58) (14, 26.53) (15, 28.42) (16, 28.9) (17, 29.61) (18, 27.88) (19, 28.92) (20, 28.09) (21, 29.12) (22, 27.38) (23, 28.51) (24, 27.85) (25, 25.42)
    }; \addlegendentry{10\%}

\addplot[
    color=green,
    mark=square*,
    ]
    coordinates {
    (1, 0.26) (2, 1.28) (3, 15.38) (4, 24.25) (5, 30.85) (6, 33.14) (7, 33.94) (8, 34.37) (9, 35.17) (10, 33.48) (11, 33.35) (12, 35.47) (13, 34.74) (14, 35.48) (15, 34.81) (16, 35.01) (17, 34.6) (18, 32.99) (19, 32.75) (20, 33.63) (21, 33.89) (22, 33.93) (23, 32.61) (24, 32.94) (25, 32.24)
    }; \addlegendentry{20\%}

\addplot[
    color=orange,
    mark=o,
    ]
    coordinates {
    (1, 15.83) (2, 31.42) (3, 33.53) (4, 35.68) (5, 36.54) (6, 36.45) (7, 37.14) (8, 37.7) (9, 37.8) (10, 37.33) (11, 38.07) (12, 38.55) (13, 37.76) (14, 38.75) (15, 38.19) (16, 38.43) (17, 38.4) (18, 37.67) (19, 38.01) (20, 37.09) (21, 37.95) (22, 37.34) (23, 35.21) (24, 37.59) (25, 37.43)
    }; \addlegendentry{40\%}

\addplot[
    color=purple,
    mark=x,
    ]
    coordinates {
    (1, 26.44) (2, 34.86) (3, 36.87) (4, 37.85) (5, 38.14) (6, 38.43) (7, 39.25) (8, 39.11) (9, 39) (10, 37.97) (11, 38.88) (12, 38.78) (13, 39.24) (14, 39.04) (15, 38.16) (16, 39.05) (17, 38.91) (18, 38.61) (19, 38.95) (20, 38.89) (21, 38.82) (22, 38.95) (23, 38.36) (24, 39.04) (25, 39.42)
    }; \addlegendentry{60\%}

\addplot[
    color=black,
    mark=+,
    ]
    coordinates {
    (1, 28.06) (2, 36.68) (3, 36.83) (4, 38.82) (5, 38.91) (6, 39.35) (7, 39.2) (8, 40.42) (9, 39.84) (10, 39.95) (11, 40.25) (12, 39.83) (13, 39.87) (14, 40.36) (15, 40.24) (16, 39.97) (17, 39.54) (18, 39.46) (19, 39.68) (20, 40.17) (21, 39.88) (22, 39.6) (23, 39.51) (24, 39.33) (25, 39.15)
    }; \addlegendentry{80\%}

\addplot[
    color=brown,
    mark=asterisk,
    ]
    coordinates {
    (1, 28.05) (2, 35.42) (3, 37.75) (4, 39.05) (5, 39.14) (6, 39.52) (7, 39.44) (8, 39.91) (9, 39.68) (10, 40.3) (11, 40.81) (12, 40.53) (13, 40.69) (14, 40.85) (15, 40.6) (16, 39.88) (17, 40.46) (18, 40.28) (19, 41.03) (20, 41.41) (21, 40.35) (22, 39.98) (23, 40.79) (24, 39.97) (25, 40.06)
    }; \addlegendentry{100\%}
\end{axis}
\end{tikzpicture}
\end{subfigure}

\caption{Hybrid testing with different amounts of real-world data mixed with synthetic NIST Juliet Test Suite vulnerability data. VulGNN accuracy and F1-score (left graph) when trained on different percentages of real-world data randomly sampled from the training set of DiverseVul, combined with a baseline of synthetic Juliet Test Suite~\cite{vulcnn2022} which is a special subset of the NIST SARD/Juliet vulnerable code collection. 
That is, 0\% is trained on a $33,360$ sample subset of the Juliet dataset, while 100\% is trained on the Juliet subset plus $419,164$ training samples from DiverseVul.
The right graph shows speed of training convergence using the DiverseVul validation set, and learning rate curves showing improvement with increasing percentage of real-world data.
}
\label{fig:sard_diversevul_exp}
\end{figure*}
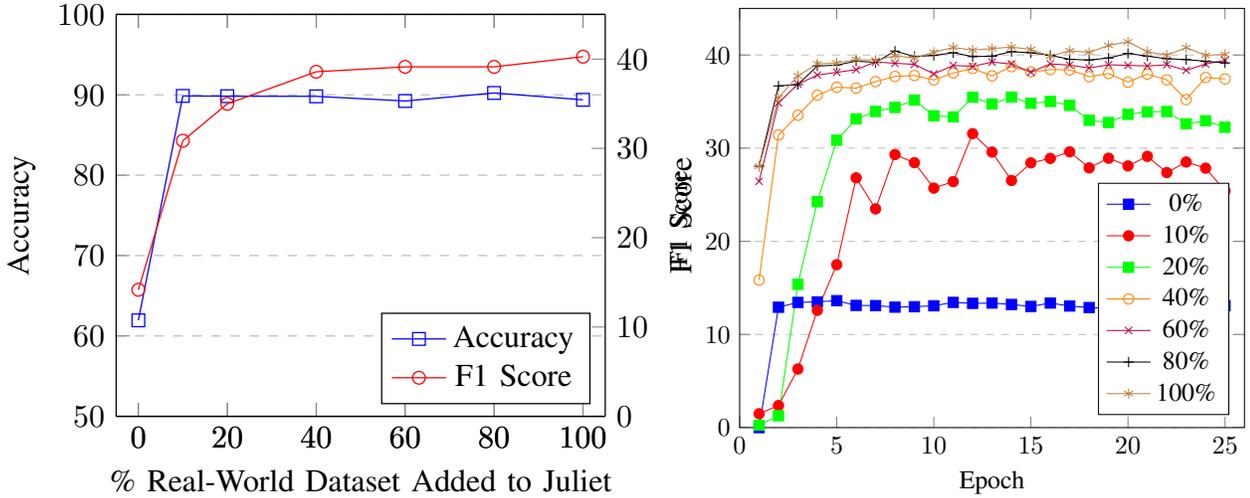

In Figure~\ref{fig:sard_diversevul_exp}, to assess the impact of real-world training data on real-world detection performance, we perform a stepped ablation study, in which we train on a baseline of synthetic data with various amounts of added real-world data.

All configurations use the entirety of our Juliet dataset as training data. Then, we create seven training sets with varying amounts of real-world data added to this synthetic base. Such data is chosen at random from the DiverseVul training set, with the amount of data selected determined by the percent label. For example, the 0\% scenario represents a training set consisting of the entirety of our Juliet dataset, but with 0\% of the DiverseVul training set added. Meanwhile, 50\% represents a training set consisting again of the entirety of our Juliet dataset, but with 50\% of the DiverseVul training set also included.

We run the experiment on seven configurations: 0\%, 10\%, 20\%, 40\%, 60\%, 80\%, and 100\% real world data included in the training set. We present the results in two formats: one that represents final performance on the DiverseVul test set after complete training, and another that shows the performance of each configuration on the DiverseVul validation set in each epoch.

\subsubsection{Test IV: Induced Training Data Balancing} \label{subsubsec:experiment_generalizability}

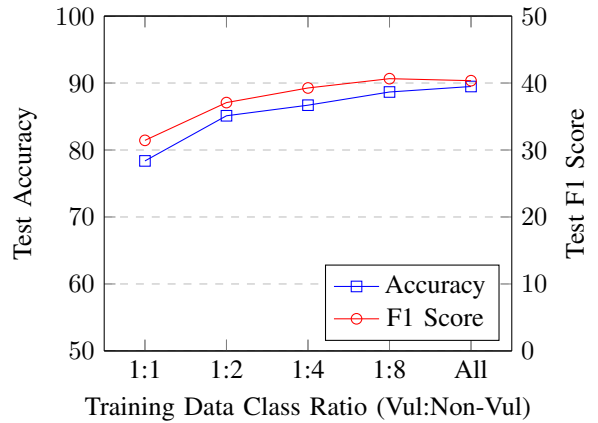
\begin{figure}
\begin{tikzpicture}
\begin{axis}[
    xlabel={Training Data Class Ratio (Vul:Non-Vul)},
    ylabel={Test Accuracy},
    ymin=50, ymax=100,
    xmin=0.5, xmax=5.5,
    xticklabels={1:1,1:2,1:4,1:8,All}, xtick={1,...,5},
    ytick={50, 60, 70, 80, 90, 100},
    ymajorgrids=true,
    grid style=dashed,
    axis y line*=left
]
\addplot[
    color=blue,
    mark=square,
    ]
    coordinates {
    (1, 78.37) (2, 85.11) (3, 86.69) (4, 88.66) (5, 89.48)
    }; \label{real_world_data_acc_plot}
\end{axis}

\begin{axis}[
    ylabel={Test F1 Score},
    ymin=0, ymax=50,
    ytick={0, 10, 20, 30, 40, 50},
    xmin=0.5, xmax=5.5,
    xticklabels={}, xtick={1,...,5},
    legend pos=south east,
    axis y line*=right
]
\addlegendimage{/pgfplots/refstyle=real_world_data_acc_plot}\addlegendentry{Accuracy}

\addplot[
    color=red,
    mark=o,
    ]
    coordinates {
    (1, 31.44) (2, 37.08) (3, 39.26) (4, 40.65) (5, 40.35)
    };
    \addlegendentry{F1 Score}
\end{axis}
\end{tikzpicture}
\caption{Accuracy and F1 Score test results on different Vul:Non-Vul training ratios}
\label{fig:vul_nonvul_ratio}
\end{figure}

The DiverseVul dataset we use in all experiments is naturally highly class-imbalanced, with an approximate 1:11.66 vulnerable to non-vulnerable sample ratio. We use class weights to counteract this feature of the dataset, but this technique is not the only option to address the imbalance. In Figure~\ref{fig:vul_nonvul_ratio} To examine the impact of this variable of the dataset on our wider results, we conduct an isolated test with induced rebalancing at various ratios. This is accomplished via random downsampling in the non-vulnerable training set, wherein some portion of the non-vulnerable training samples are randomly chosen and discarded such that the remaining non-vulnerable training samples form some desired ratio with the vulnerable training samples (e.g. 1:2). Following this procedure, we retain the imbalance in the test dataset, which is more representative of real-world data (only a minority of functions are vulnerable in reality), while still creating a more balanced training dataset. We continue to use class weights in all scenarios. It is already known that without this feature, any dataset imbalance greatly negatively impacts the F1 score.

\subsection{Results}

\begin{table*} \centering
\bgroup
\setlength\tabcolsep{0.33cm}
\def\arraystretch{1.8}
\caption{Vulnerability detection performance of deep learning models for two scenarios. The best performing result in each column is shown in bold. Results in the \textit{``Train on Prev+Diverse, Test on Unseen Projects''} column represent results from the experiment described in Section~\ref{subsec:experiment_generalizability}. Results in the \textit{``Train \& Test on Prev+Diverse''} column represent results from the experiment described in Section~\ref{subsec:experiment_prev_plus_div}. All results, except for VulGNN, are from DiverseVul~\cite{chen2023diversevul} (see their Tables 4 and 5) (test on same and unseen data respectively; marked with $^*$), and Table 6 (optimized hyperparameters; marked with $^\dag$). See the aforementioned sections for further details on dataset split and experiment setup.
}
\label{tab:diverse_comparison}
\begin{tabular}{c|c|ccccc|ccccc}
\hline \hline
\multirow{2}{*}{\textbf{Family}} & \multicolumn{1}{c|}{\multirow{2}{*}{\textbf{Architecture}}} & \multicolumn{5}{c|}{\textbf{Train on Prev+Diverse, Test on Unseen Projects}} & \multicolumn{5}{c}{\textbf{Train (80\%) \& Test (10\%) on Prev+Diverse} (no overlap)}   \\ \cline{3-12} 
                        & \multicolumn{1}{c|}{}                                 & F1             & Acc            & Prec           & Rec            & FPR            & F1             & Acc            & Prec           & Rec            & FPR            \\ \hline \hline
\multirow{2}{*}{GNN}    & ReVeal$^*$                                            & 8.67           & 85.88          & 5.67           & 18.46          & 11.58          & 29.76          & 82.81          & 23.75          & 39.83          & 12.87          \\ \cline{2-12} 
                        & \textbf{VulGNN (Ours)}                                    & \textbf{18.17} & 93.17      & 16.18          & 20.72          & 4.08           & 40.35          & 89.48          & 36.84          & 44.6           & 6.63           \\ \hline \hline
\multirow{4}{*}{BERT}   & RoBERTa*                                              & 4.40           & 95.59          & 10.46          & 2.78           & 0.90           & 34.98          & 91.68          & 46.02          & 28.22          & 2.85           \\ \cline{2-12} 
                        & CodeBERT$^*$                                          & 11.94          & 94.19          & 13.34          & 10.80          & 2.65           & 37.85          & 90.48          & 39.25          & 36.54          & 4.87           \\ \cline{2-12} 
                        & CodeBERT$^\dag$                                       & 14.74          & 92.16          & 12.21          & 18.60          & 5.06           & 41.72          & 89.39          & 36.97          & 47.89          & 7.04           \\ \cline{2-12} 
                        & GraphCodeBERT$^*$                                     & 9.25           & 94.74          & 12.48          & 7.35           & 1.95           & 36.79          & 90.32          & 38.18          & 35.51          & 4.96           \\ \hline \hline
\multirow{4}{*}{GPT-2}  & GPT-2 Base$^*$                                        & 6.02           & 95.06          & 9.82           & 4.34           & 1.51           & 33.03          & 91.73          & 46.18          & 25.71          & 2.58           \\ \cline{2-12} 
                        & CodeGPT$^*$                                           & 7.72           & 94.47          & 9.86           & 6.35           & 2.19           & 35.23          & 91.36          & 43.48          & 29.62          & 3.32           \\ \cline{2-12} 
                        & PolyCoder$^*$                                         & 11.39          & 92.73          & 10.25          & 12.81          & 4.24           & 31.96          & 91.97          & 48.76          & 23.78          & \textbf{2.15}  \\ \cline{2-12}
                        & PolyCoder$^\dag$                                      & 13.63          & 89.76          & 9.84           & \textbf{22.16} & 7.68           & 36.68          & 86.48          & 29.19          & 49.36          & 10.32          \\ \hline \hline 
\multirow{5}{*}{T5}     & T5 Base*                                              & 9.73           & \textbf{96.16} & \textbf{34.00} & 5.68           & \textbf{0.42}  & 42.33          & 91.96          & 49.14          & 37.17          & 3.32           \\ \cline{2-12} 
                        & CodeT5 Small$^*$                                      & 9.39           & 94.91          & 13.35          & 7.24           & 1.78           & 45.10          & 91.85          & 48.41          & 42.22          & 3.88           \\ \cline{2-12}
                        & CodeT5 Small$^\dag$                                   & 17.21          & 93.87          & 16.95          & 17.48          & 3.24           & \textbf{48.28} & 89.57          & 39.80          & \textbf{61.33} & 7.99           \\ \cline{2-12} 
                        & CodeT5 Base$^*$                                       & 9.14           & 95.56          & 18.03          & 6.12           & 1.05           & 45.69          & 92.11          & 50.36          & 41.81          & 3.55           \\ \cline{2-12} 
                        & NatGen$^*$                                            & 9.30           & 95.49          & 17.38          & 6.35           & 1.14           & 47.15          & \textbf{92.30} & \textbf{51.81} & 43.25           & 3.47          \\ \hline \hline
\end{tabular}
\egroup

\end{table*}

The results of tests \textit{I} and \textit{II} are shown in Table \ref{tab:diverse_comparison}. We include the performance data of models tested in the DiverseVul paper for comparison convenience. When sorting by F1-score, VulGNN leads any other detector by about 6\% in unseen project performance, achieving this improvement with a minor expense to accuracy.

\begin{table} 
\caption{The number of learnable weight parameters for each network.}
\centering
\bgroup
\setlength\tabcolsep{0.45cm}
\def\arraystretch{1.8}
\begin{tabular}{c|c|c}
\hline \hline
Family                 & Architecture & Number of Parameters \\ \hline \hline
\multirow{2}{*}{GNN}   & ReVeal       & 1.28 Million         \\ \cline{2-3} 
                       & VulGNN       & 1.10 Million        \\ \hline \hline
\multirow{3}{*}{BERT}  & RoBERTa      & 125 Million          \\ \cline{2-3} 
                       & CodeBERT     & 125 Million          \\ \cline{2-3} 
                       & GraphCodeBERT    & 125 Million          \\ \hline \hline
\multirow{3}{*}{GPT-2} & GPT-2 Base   & 117 Million          \\ \cline{2-3} 
                       & CodeGPT      & 124 Million          \\ \cline{2-3} 
                       & PolyCoder    & 160 Million          \\ \hline \hline
\multirow{4}{*}{T5}    & T5 Base      & 220 Million          \\ \cline{2-3} 
                       & CodeT5 Small & 60 Million           \\ \cline{2-3} 
                       & CodeT5 Base  & 220 Million          \\ \cline{2-3} 
                       & NatGen       & 220 Million          \\ \hline \hline
\end{tabular}
\egroup

\label{tab:model_size_comparison}
\end{table}

It is also important to interpret these results in the context of model size. Table \ref{tab:model_size_comparison} shows the number of parameters in each model. Note that graph-based detectors also require significant preprocessing steps to generate graph data from source code, so the number of parameters cannot be used as a direct efficiency comparison. However, the graphs used by VulGNN are often generated as a byproduct of other development tools that conduct static analysis, so this overhead may be negligible in real world projects where these artifacts can be reused.


\section{Discussion} \label{sec:discuss}

\subsection{Discussion of Results} \label{subsec:results_discussion}


Our work culminates in a detection system that, per our test findings, is competitive with state-of-the-art systems in both unseen and in-distribution evaluations. In the Unseen Project configuration of DiverseVul, VulGNN improves F1 by approximately 6\% compared to the best baseline, while maintaining competitive accuracy. Although LLM-based models achieve higher raw accuracy in some settings, they often suffer from poor recall and huge parameter counts.

In terms of efficiency, VulGNN has only 1.1M parameters compared to 60M--220M in LLMs. This is two orders-of-magnitude reduction in memory footprint and compute cost. Even without precise inference benchmarks, it is evident that VulGNN can be deployed on commodity GPUs and potentially CPUs within CI/CD pipelines, while LLMs require dedicated high-memory GPUs. Thus, VulGNN balances detection capability with practical deployability.

\subsection{Threats to Validity} \label{subsec:threats_to_validity}

\subsubsection{Internal Threats to Validity}
Internal validity refers to whether the observed outcomes in our study are genuinely caused by the factors under investigation rather than hidden confounders. To reduce such risks, we fully automated the data preparation, preprocessing, and evaluation pipelines, ensuring consistent treatment across all experimental runs. 

However, threats remain. In particular, parameter configurations and implementation choices (e.g., preprocessing heuristics, splitting strategies, or default hyperparameters of tools) may have inadvertently influenced the results. Another potential source of bias lies in the selection of subsets of the dataset for evaluation, which could favor certain vulnerability types over others. We mitigated this by using multiple runs, sampling randomization, and validation on independent splits, but confounding influences cannot be entirely ruled out.

\subsubsection{External Threats to Validity}
External validity concerns the degree to which our results generalize beyond the studied context. While we relied on DiverseVul, which is one of the largest and most carefully curated datasets currently available, the findings may not directly generalize to all software domains or to industrial projects. Real-world code bases often differ in scale, coding style, testing practices, and developer workflows, and such differences can affect how vulnerabilities manifest and how models perform in practice.

Moreover, vulnerability datasets continue to evolve. Even though DiverseVul contains a larger number of instances with higher label correctness (validated by random sampling) than previous datasets, it is still subject to inherent limitations of any benchmark dataset. For instance, it may not fully represent emerging vulnerability types, non-C/C++ projects, or industrial code developed under proprietary processes. Replications in additional settings, particularly on industrial datasets or longitudinal corpora, would be valuable to assess the generalizability of our findings.

To further mitigate threats of project-specific bias, we conducted two complementary evaluation setups: (1) training on Prev+Diverse and testing on entirely unseen projects, and (2) training and testing on Prev+Diverse with no overlaps. The unseen-project setting provides stronger evidence of generalizability, as models are not exposed to project-specific coding styles or artifacts during training. While this design reduces the risk of overfitting to particular projects, it still does not guarantee transferability to industrial contexts, where code bases may differ substantially in complexity, scale, or development practices.

\subsubsection{Consruct Threats to Validity}
Construct validity relates to whether the measures and datasets used in the study adequately capture the theoretical concepts of interest. Our study relied on publicly available vulnerability datasets. However, prior work (including DiverseVul itself) has highlighted significant limitations in these resources: for example, incorrect labels, incomplete mappings of vulnerabilities that span across multiple functions, and inconsistencies across datasets such as BigVul~\cite{bigvul}, CVEFixes~\cite{cvefixes}, and CrossVul~\cite{crossvul}. By adopting DiverseVul, we aimed to mitigate these issues, since it provides more data and a higher rate of correct labeling (based on manual inspection of a random sample).

At the same time, our decision not to rely exclusively on synthetic datasets such as SARD/Juliet introduces another trade-off. While SARD/Juliet provides more reliably correct labels, its data consists of small, often artificial code samples that are less representative of real-world software, limiting ecological validity. Thus, while our dataset choice strengthens the realism of the study, there remains a residual threat that the chosen benchmarks may still not perfectly capture the construct of ``real-world vulnerability detection.''

\subsubsection{Conclusion Threats to Validity}
Conclusion validity concerns whether the conclusions we draw from the observed results are justified given the evaluation measures. Our analysis relied on commonly used performance metrics (precision, recall, F1-score, accuracy, and false positive rate), which provide complementary views of model behavior. Using multiple measures reduces the risk of drawing misleading conclusions based on a single metric (e.g., high accuracy in imbalanced data).

However, threats remain. First, while these metrics quantify performance, they may not fully reflect the practical utility of a vulnerability detection system in real-world development. For instance, a low false positive rate may still translate into a large number of alerts in large projects, potentially reducing tool adoption despite favorable metrics. Second, results may be sensitive to dataset imbalance: metrics such as accuracy can be inflated when non-vulnerable instances dominate, while F1-score may emphasize trade-offs between precision and recall but overlook costs associated with misclassifications. Finally, since we did not perform statistical tests, we cannot formally assess whether observed differences between models are significant or due to chance.

To mitigate the limitations in metric representation, we reported multiple metrics to provide a holistic perspective, and we interpreted results cautiously, emphasizing observed patterns and relative performance rather than absolute claims of superiority. To minimize the impact of random data sampling and parameter initialization, every configuration in each experiment is trained and tested three times, with the presented test results being the average over the three runs. Nonetheless, we acknowledge that further studies with larger samples, alternative evaluation frameworks, or statistical testing would be needed to strengthen the robustness of our conclusions.

Another consideration is that results varied across the two evaluation designs. The Train on Prev+Diverse, Test on Unseen Projects setting provides stronger evidence of generalizability, as the model is not exposed to project-specific artifacts during training, while the Train and Test on Prev+Diverse (with no overlaps) setting reflects overall performance on the combined dataset. This difference highlights that conclusions must be interpreted within the context of the evaluation setup, and that replications with additional testing scenarios would further strengthen confidence in the findings.

\section{Related Work} \label{sec:related}

\begin{table*}[h!]\footnotesize
\centering
\caption{Comparison of VulGNN and Representative recent GNN-Based Vulnerability Detectors. 
}
\begin{tabular}{p{0.11\linewidth} | p{0.035\linewidth} | p{0.105\linewidth}| p{0.2\linewidth}| p{0.1\linewidth}| p{0.2\linewidth}| p{0.05\linewidth}}

\hline
\textbf{Model} & \textbf{Year} & \textbf{Graph / Input Representation} & \textbf{Model Architecture / Key Innovation} & \textbf{Languages / Scope} & \textbf{Evaluation Dataset(s)} & \textbf{Training Code Availability} \\ 
\hline
\textbf{VulGNN (Ours)} & 2025 & CPG / AST+CFG+PDG & GNN design emphasizing generalization, regularization, cross-project robustness & C/C++ & SARD/Juliet + DiverseVul (Real-world) & \cmark \\
\hline
ReVeal~\cite{chakraborty2021deep} & 2022 & CPG & Gated Graph Neural Network (GGNN) with SMOTE and triplet loss & C/C++ & Devign~\cite{zhou2019devign} + vulnerable samples collected from Chrome and Debian projects & \cmark \\

\hline
BGNN4VD~\cite{cao2021bgnn4vd} & 2021 & Code graph (bidirectional edges) & Bidirectional GNN integrating forward & backward flows & C/C++ \& NVD / GitHub projects & \xmark \\

\hline
ReGVD~\cite{nguyen2022regvd} & 2022 & Token-level graph over code tokens & Lightweight GNN with residual layers, mixed pooling & Multi-language (incl. C/C++) & CodeXGLUE vulnerability benchmark & \cmark \\ 

\hline
LineVD~\cite{hin2022linevd} & 2022 & PDG at statement granularity & Hybrid Transformer + GAT — statement-level classification & C/C++ & Real vulnerability projects (Linux, Chromium) & \cmark \\

\hline
MVD~\cite{cao2022mvd} & 2022 & Flow-sensitive PDG (statement-level) & Flow-sensitive GNN (FS-GNN) tailored for memory errors & C/C++(memory bugs) & Memory-related vulnerabilities (4,353 samples) & \xmark \\

\hline
TACSan~\cite{zeng2024tacsan} & 2024 & Graph over TAC (3-address code) & GNN over normalized intermediate representation & C/C++ & SARD + multiple CWE datasets & \xmark \\

\hline
Joint Graph + Transformer~\cite{jin2025source} & 2025 & Merged multi-graph (AST, CFG, PDG, etc.) & GAT + Transformer fusion, pre/post fusion & C/C++ & SARD + ``Real-Vul'' dataset & \xmark \\

\hline
ExplainVulD~\cite{haque2025explainable} & 2025 & CPG (AST+CFG+DFG) & Edge-aware GATv2 + dual-channel embeddings + explanation module & C/C++ & ReVeal dataset & \xmark \\

\hline
DSHGT~\cite{zhang2024dshgt} & 2024 & Heterogeneous CPG (typed nodes/edges) & Heterogeneous graph transformer, dual-supervisor architecture & C/C++ (and cross-language) & Real-world projects + transfer tests & \xmark \\

\hline
VulTriNet~\cite{yang2025vultrinet} & 2025 & Intermediate representation (IR) + code graph & GNN on IR + noise-mitigation & C/C++ & (not fully detailed) & \xmark \\
\hline
\end{tabular}
\label{tab:related}
\end{table*}

One of the first GNN-based methods to detect code vulnerabilities was presented by Zhou et. al \cite{zhou2019devign}, which was introduced as \textit{Devign}  that learned from AST, CFG, and DFG code graph features to detect code vulnerabilities on open-source C projects. Cheng et al. \cite{cheng2021deepwukong} used GNNs to embed code fragments, preserving control-flow, data-flow, and natural language information in their \textit{XFG} representation (program slices). Unlike VulGNN, Cheng et al. \cite{cheng2021deepwukong} used program slicing on PDGs, which is a code graph representation that is commonly used to only contain relevant code entities (nodes and edges) that are related to the vulnerability. Their GNN is based on two graph convolutional layers, while we used a different convolutional layer called GeneralConv with five layers. In addition, Cheng et al. \cite{cheng2021deepwukong} worked on a semi-synthetic and trained on only 10 CWE classes, while we have trained on semi-synthetic and real-world vulnerabilities that included 150 CWE classes. Furthermore, they used Doc2Vec's tokenizer \cite{le2014distributed} to embed their code, while we used StarCoder.

Some other early GNN-based methods demonstrated the benefit of modeling code structure, but often struggled with scalability and generalization. BGNN4VD~\cite{cao2021bgnn4vd} improved over static analyzers by using bidirectional edges, while ReGVD~\cite{nguyen2022regvd} showed that even lightweight token-level GNNs could achieve competitive results on CodeXGLUE~\cite{lu2021codexglue}. However, both approaches were evaluated mainly within single datasets, making their robustness under cross-project settings less clear. In contrast, VulGNN explicitly targets generalization across large-scale, real-world C/C++ projects, prioritizing stability under domain shifts rather than only within-dataset performance.

Later works introduced more sophisticated graph representations or hybrid architectures. LineVD~\cite{hin2022linevd} and MVD~\cite{cao2022mvd} refined statement-level program dependence graphs for detecting memory vulnerabilities, and TACSan~\cite{zeng2024tacsan} normalized code into intermediate representations to reduce noise. While effective, these models were tailored to specific vulnerability types or relied heavily on curated datasets. By comparison, VulGNN is designed as a general-purpose detector, maintaining consistent performance across varied vulnerability classes and projects without overfitting to one representation or benchmark.

More recent advances emphasize either performance boosts through model complexity or explainability. For instance, Joint Graph + Transformer~\cite{jin2025source} and DSHGT~\cite{zhang2024dshgt} integrate transformers or heterogeneous graph modeling to achieve higher accuracy, while ExplainVulD~\cite{haque2025explainable} and VulPathFinder~\cite{atashin2025learning} provide interpretability by surfacing influential nodes or paths. Other works such as VulTriNet~\cite{yang2025vultrinet} continue refining graph construction and noise handling. These systems highlight promising directions but often involve heavier models or focus narrowly on explanation. VulGNN, by contrast, strikes a balance between expressiveness and efficiency, aiming for reproducible, cross-project robustness with lower complexity, which makes it more deployable in practical vulnerability detection pipelines.

Finally, in Table~\ref{tab:related} we show the differences between our proposed model and other GNN models. We did not directly benchmark VulGNN against every prior GNN-based vulnerability detector included in the related work due to significant reproducibility and comparability limitations. Among the cited studies, only two (ReGVD~\cite{nguyen2022regvd} and LineVD~\cite{hin2022linevd}) provide publicly available training code, making them the only reproducible GNN baselines identified. However, both rely on large pre-trained transformer backbones (e.g., GraphCodeBERT and CodeBERT) exceeding 125 million learnable parameters, resulting in models that are orders of magnitude larger and computationally heavier than VulGNN. Additionally, the LineVD~\cite{hin2022linevd} is a statement-level vulnerability detection model (node classification), while VulGNN is a function-level vulnerability detection (graph classification) model. In other words, our model and LineVD approach the vulnerability detection problem at different granularities of code, which makes them unfit to compare. In contrast, our model is intentionally designed to be lightweight and efficient while maintaining competitive predictive capability. Furthermore, many other GNN-based detectors employ different code representations (e.g., program dependence graphs or intermediate representations), target languages other than C/C++, or address tasks such as statement-level or multi-class vulnerability classification, which are conditions incompatible with our function-level binary classification setup. Given these disparities, integrating those approaches into our experimental framework would require non-trivial re-engineering and could lead to unfair or misleading comparisons. Consequently, our evaluation prioritizes generalization, reproducibility, and deployment feasibility, which are key goals that distinguish VulGNN from previous GNN architectures. Table~\ref{tab:related} already highlights the architectural and dataset-level differences across these studies, underscoring that our decision reflects a deliberate design focus rather than an omission of comparable baselines.

\section{Conclusions and Future Work} \label{sec:conclusion}

VulGNN is a lightweight graph neural network based deep architecture for code vulnerability detection that outperforms the state-of-the-art ReVeal GNN and is comparable to large language models, but with two orders of magnitude fewer weights. We show that incorporating even a small amount of real-world vulnerable code examples ($\sim$42K samples or 10\% of DiverseVul) to balance the purely synthetic NIST Juliet dataset, significantly improves accuracy from 62\% to 90\%.
F1 performance continues improving from 31\% to almost 40\% as the amount of real-world DiverseVul samples increases from 10\% to 40\% ($\sim$167.7K DiverseVul samples).
The lightweight nature of VulGNN makes it significantly more practical to include in code development environments, such as continuous integration (CI) pipelines, and could prove a powerful tool in combination with emerging AI code reviewers.

Future work primarily involves examining other model variations, training techniques, and generalization across programming languages. Variations of particular interest include using different convolutional layers, using a pre-trained ``foundation'' embedding model to create node representations, and introducing data augmentation for enhanced network communicativity.


\section{Acknowledgments}

This work was partially supported by the MizzouForward Undergraduate Research Training Grant.

This work was partially supported by the U.S. National Science Foundation award 2243619 and Army Engineer Research and Development Center - Information Technology Laboratory (ERDC-ITL) contract W912HZ23C0041.

\bibliographystyle{IEEEtran}
\bibliography{biblio}

\begin{IEEEbiography}[{\includegraphics[width=1in,height=1.25in,clip,keepaspectratio]{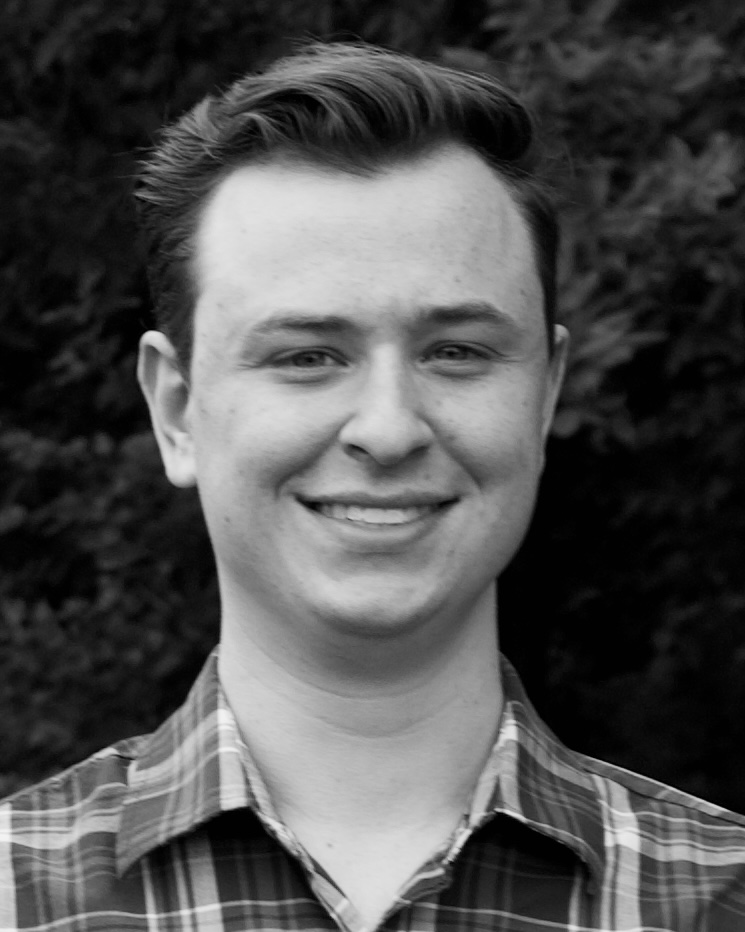}}]{Miles Farmer}
is an undergraduate computer science and mathematics student at the University of Missouri, where he is a member of the Computational Imaging and Visualization Analysis (CIVA) Laboratory. His current research interests include graph-based deep learning and its applications to software analysis and security, deep learning and optimization techniques, and language processing.
\end{IEEEbiography}

\begin{IEEEbiography}[{\includegraphics[width=1in,height=1.25in,clip,keepaspectratio]{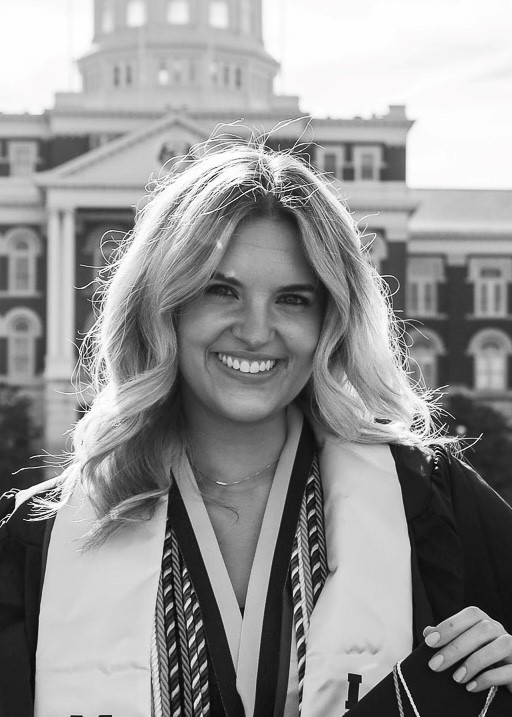}}]{Anne Watson} graduated from the University of Missouri with a B.S. in Electrical Engineering and a B.S. in Economics. During her time at the University of Missouri, she was a member of the Computational Imaging and Visualization Analysis (CIVA) Lab. Her research interests include Convolutional Neural Networks and their applications to software vulnerabilities. 

\end{IEEEbiography}

\begin{IEEEbiography}[{\includegraphics[width=1in,height=1.25in,clip,keepaspectratio]{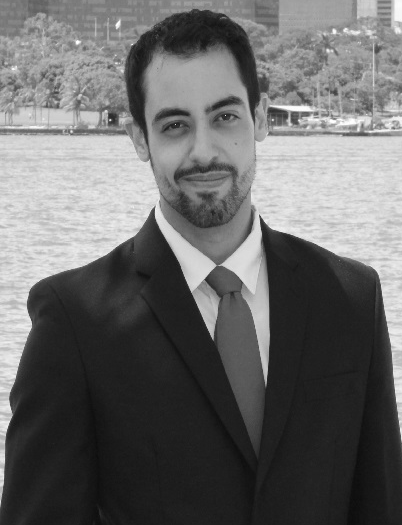}}]{Hialo Muniz Carvalho} received the B.S degree in Software Engineering from the University of Brasilia, Brazil,  in 2014. Currently working as a Guest Researcher at the National Institute of Standards and Technology with the Software Assurance Metrics And Tool Evaluation (SAMATE) group, his current research is focused on developing machine learning models for vulnerability detection in source code. His current research interests include Graph Neural Networks applied to several areas, including software analysis and software security.

\end{IEEEbiography}
\begin{IEEEbiography}[{\includegraphics[width=1in,height=1.25in,clip,keepaspectratio]{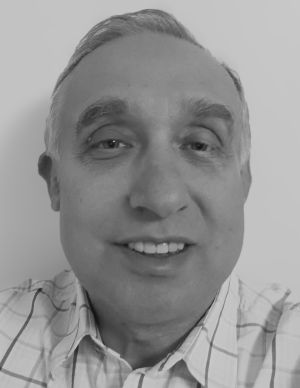}}]{Vadim Okun}(Member, IEEE) is a Computer Scientist at the National Institute of Standards and Technology, where he is leading the Software Assurance Metrics And Tool Evaluation (SAMATE) team. His current research focuses on software assurance, in particular, the effect of tools on security. Previously, he contributed to the development of automated software testing methods: specification-based mutation testing and combinatorial testing. He received a Ph.D. degree in Computer Science from the University of Maryland Baltimore County.

\end{IEEEbiography}
\begin{IEEEbiography}[{\includegraphics[width=1in,height=1.25in,clip,keepaspectratio]{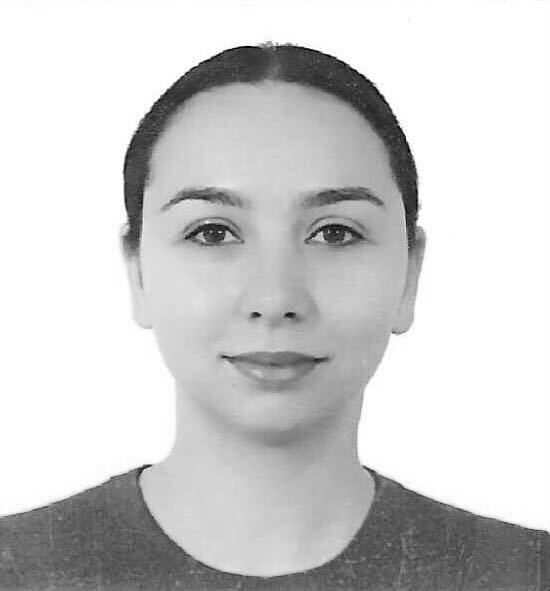}}]{Zineb Maasaoui} received the master degree in mathematics and mechanics from ENSEIRB-MATMECA Bordeaux, in France in 2017. Currently a PhD student at the university of Grenoble, France she is also a guest researcher at the National institute of standards and technology with the Software Assurance Metrics And Tool Evaluation group. Her research is focused on datasets evaluation and improvement. Her main domain of interests are AI and software security.

\end{IEEEbiography}

\begin{IEEEbiography}[{\includegraphics[width=1in,height=1.25in,clip,keepaspectratio]{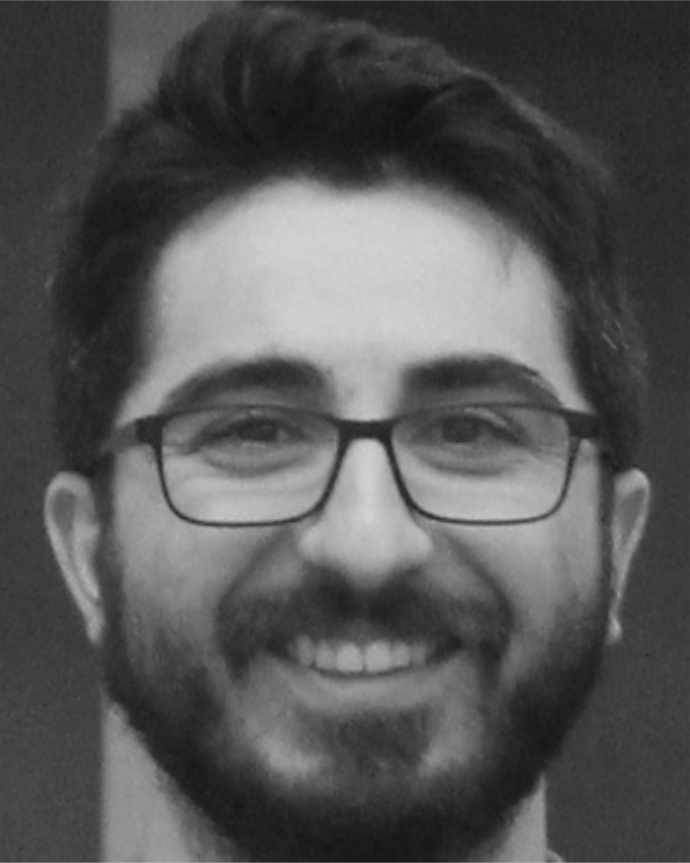}}]{Ekincan Ufuktepe}
(Member, IEEE) received the B.S. degree in computer engineering from the Izmir University of Economics, in 2011, and the M.S. and Ph.D. degrees in computer engineering from the İzmir Institute of Technology, in 2014 and 2019, respectively. In 2013, he was a Research Intern with the Security and Trust Department, SAP Laboratories France. From 2019 to 2021, he was a Postdoctoral Researcher at the Computational Imaging and Visualization Analysis (CIVA) Laboratory at the University of Missouri-Columbia. He is currently an Associate Teaching Professor at the University of Missouri-Columbia Electrical Engineering and Computer Science Department. His current research interests include software testing, program analysis, and software security.
\end{IEEEbiography}

\begin{IEEEbiography}[{\includegraphics[width=1in,height=1.25in,clip,keepaspectratio]{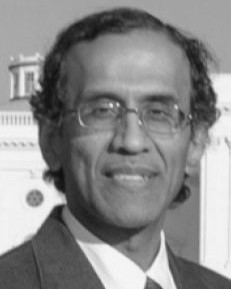}}]{Kannappan Palaniappan}
(Senior Member, IEEE) received the Ph.D. degree from the University of Illinois at Urbana–Champaign, Champaign, IL, USA. He is currently a Curator's Distinguished Professor with the Electrical Engineering and Computer Science Department, University of Missouri. His research interests include the synergistic intersection of image and video processing, computer vision, high-performance computing, and artificial intelligence to understand, quantify, and model physical and computational systems.
\end{IEEEbiography}

\end{document}